\documentstyle[aaspp4,12pt,amsmath]{article}

\lefthead{FRYER \& Heger} 

\righthead{Collapsar Engines from Binary Mergers}

\begin{document} 

%\begin{center}

%\end{center}

%\vspace{1.cm}

\title{Binary Merger Progenitors for Gamma-ray
  Bursts and Hypernovae} 
\author{Chris L. Fryer\footnote{Also at:  Physics Department, University of
    Arizona, Tucson, AZ, 85721} and Alexander Heger\footnote{Also at: 
    Enrico Fermi Institure, University of Chicago, 
    5640 S. Ellis Ave, Chicago, IL 60637}} 
\affil{Theoretical Astrophysics, Los Alamos National Laboratories, \\ 
  Los Alamos, NM 87545} 
\authoremail{fryer@lanl.gov, aheger@lanl.gov}

\begin{abstract}
  
  The collapsar model, the now leading model for the engine behind
  gamma-ray bursts and hypernovae, requires that a star collapses to
  form a black hole surrounded by an accretion disk of high-angular
  momentum material.  The current best theoretical stellar models,
  however, do not retain enough angular momentum in the core of the
  star to make a centrifugally supported disk.  In this paper, we
  present the first calculations of the helium-star/helium-star merger
  progenitors for the collapsar model.  These progenitors invoke the
  merger of two helium cores during the common envelope inspiral phase
  of a binary system.  We find that, in some cases, the merger can
  produce cores that are rotating 3-10 times faster than single stars.
  He-star/He-star gamma-ray burst progenitors have a very different
  redshift distribution than their single-star gamma-ray burst
  progenitors and we discuss how gamma-ray burst observations can
  constrain these progenitors.
  
\end{abstract}


\section{Introduction}
\label{sec:intro}

The accurate localizations of long-duration gamma-ray bursts (GRBs)
have led to an increasing set of data indicating that these phenomena are
associated with the deaths of massive stars (see Zhang et al.\ 2004
for a review).  Some of the most convincing evidence is the
simultaneous (both spatially and temporally) occurrence of a GRB (GRB
030329) with a bright, energetic Type Ic supernova, SN 2003dh (Price
et al.\ 2003; Hjorth et al.\ 2003; Stanek et al.\ 2003).  At the same
time, a class of supernovae, characterized by strong explosions and
possibly large asymmetries (Maeda et al.\ 2003) and a set of X-ray
flashes with weak gamma-ray signals (e.g., Fynbo et al.\ 2004) were
both discovered in the data.

These explosive phenomena have many similarities and are generally
grouped into a large class termed ``hypernovae'' with GRBs making up a
subset of this class.  It has been argued that non-GRB hypernovae are
GRBs not directed along our line-of-sight or GRB-like jet explosions
where the jet is no longer relativistic when, and if, it breaks out of
the star (e.g., MacFadyen et al.\ 2001; Fynbo et al.\ 2004; Zhang et
al.\ 2004).  The collapsar model (Woosley 1993, MacFadyen \& Woosley
1999) is gradually becoming the favored engine behind all types 
of hypernova explosions.

The collapsar model can not explain all supernova observations (Fryer
et al.\ 1999), and there is growing observational evidence showing
that these explosions are rare.  Radio observations suggest
that at most 5\% of all Type Ib/Ic supernovae can be produced in GRBs
(Berger et al. 2003).  Likewise, optical observations (correcting for
observational biases) of hypernovae suggest that this fraction is less
than 1\% (Podsiadlowski et al. 2004).  Because the event rate is so
small, the progenitor evolution can be much more exotic.  Here we
focus on progenitors for the collapsar engine alone.

The collapsar model is part of a class of models invoking accretion
disks around black holes (Popham et al.\ 1999; Fryer et al.\ 1999).
In the collapsar mechanism, the black hole is formed when
the collapse of a massive star fails to produce a strong supernova
explosion, leading to the ultimate collapse into a black hole.  If the
stellar material falling back and accreting onto the black hole has
sufficient angular momentum, it can hang up, forming a disk.  This
disk, by neutrino annihilation or magnetic fields, is thought to
produce the jet which finally results in a GRB or a hypernova that
we observe (Popham et al. 1999; MacFadyen \& Woosley 1999).  For such
a mechanism to work, the star must satisfy three criteria:
\begin{itemize}
\item[\textbf{I:}] The star must collapse to a black hole.  This can
  occur in stars that \emph{initially} produce weak or no explosions.
\item[\textbf{II:}] The star must have sufficient angular momentum to
  form a disk around that black hole.  The ideal range of angular
  momentum, $j$, in the core lies between $10^{16} < j < 10^{18} {\rm
    cm^2 \, s^{-1}}$.
\item[\textbf{III:}] The star must lose its hydrogen envelope.  This
  criterion is necessary for the jet to remain relativistic (Zhang et
  al.\ 2004).  Some hypernovae may not need to satisfy this criterion.
\end{itemize}

In principal, massive single stars can satisfy these criteria.  In the
absence of winds, massive stars above $\sim 18-25$M$_\odot$ are all
believed to collapse to form black holes (Fryer 1999).  With the
inclusion of mass loss from winds, binary interactions, or both, many
massive stars will still collapse to form black holes (Fryer et al.\
2002; Heger et al.\ 2003).  A sizable fraction of single stars, easily
enough to produce the rates required to form hypernovae and gamma-ray
bursts, will collapse after losing their entire hydrogen envelope.  If
single stars are the dominant GRB progenitor, the GRB rate should 
decrease dramatically at high redshift where winds don't eject the 
hydrogen envelope as effictively.  In general, single stars that
do lose their hydrogen envelope also lose a lot of their angular 
momentum, so it is not clear that single stars can match all three 
criteria.

To solve this angular momentum deficiency, in particular caused by the
angular momentum loss from stellar winds, Fryer et al.\ (1999)
proposed that the cores of massive stars in binaries would merge in a
common envelope phase, leaving behind a merged core and ejecting much
of the hydrogen envelope without slowing down the rotation of the
core.  But what kind of binary scenario will cause the merged core to
rotate rapidly?  Figure 1 shows two evolutionary paths of binaries
that lead to a merged core.  In the first path (I), the primary star
engulfs its companion and the two stars merge while the companion is
still on the main sequence.  This scenario is akin to the leading
scenario for the progenitor of supernova 1987A (Podsiadlowski 1992), 
which can spin up the outer part of the core (Ivanova et al. 2002).
Alternatively, the stars may not merge in the first common envelope
phase.  If the companion has nearly the same mass as its primary, it
may also evolve off the main sequence before the primary collapses,
leading to a second common envelope phase.  If these two helium stars
merge (II), considerable angular momentum would be deposited into the
merged core.  Fryer et al.\ (1999) termed these mergers the He-merger
formation scenario (merger of two helium cores) for collapsars and it
is these objects that we study in this paper.

How do these objects fit into the grand scheme of binary populations?
Figure 2 shows a chart of the fates of massive stars.  X-ray binaries
are formed in systems where the two stars do not merge.  The
progenitor of SN 1987A may have formed from a binary of a massive star
and a low mass companion that merged and retained much of its hydrogen
envelope.  Although it may seem unlikely that many binaries consisting
of two stars of nearly the same mass exist, these binaries may be the
primary scenario for forming double neutron star systems (Brown 1995;
Fryer et al.\ 1999; Belczynski et al.\ 2002).  Such incidences may be
rare, but they easily can form the observed hypernova population
(Belczynski et al. 2002 found that this rate could exceed 10 mergers
per Myr).

In this paper, we present the first in a series of simulations
studying the actual merger of helium cores to determine whether such a
merger phase can produce a collapsar, and ultimately a hypernova or
gamma-ray burst.  In \S 2, we describe the combination of codes used
to model the merger and stellar evolution of these binary systems.
The results of these calculations are given in \S 3, the fate of the
merged cores is discussed in \S 4, and their implications for
gamma-ray bursts are given in \S 5.

\section{Computations}

Determining whether binary mergers (of the He-He merger class) play a
role in collapsars requires a range of physics and numerical
techniques including implicit hydro codes capable of modeling the
evolution of a star through its entire life and simulations following
the multi-dimensional effects of the merger itself.  As such, this
study, by necessity, is a multi-step process:
\begin{itemize}
\item{\bf I:} Evolve each component star of the binary system from
  birth to the point that the common envelope phase, and hence the
  merger, begins (e.g., after the stars have moved off the main
  sequence).  This provides the structure of the stars for the next
  step.
\item{\bf II:} Model the actual merger process as the two stars evolve
  through a common envelope to determine the thermodynamic and
  composition structure and angular momentum distribution of the
  merged core for the next step.
\item{\bf III:} Evolve the merged star through the rest of its life to
  the collapse of its iron core.  These simulations will produce the
  detailed structure and angular momentum profiles with which we can
  compare these binary collapsar progenitors with their single star
  counterparts.
\end{itemize}
Steps I and III require full stellar evolution codes, and for these
steps we use a modified version of the stellar evolution code KEPLER
(Weaver et al.\ 1978; Heger et al.\ 2000; Woosley et al. 2002).  The
actual merger process (Step II) requires multi-dimensional
hydrodynamic calculations, and for this step we use the 3-dimensional
smooth particle hydrodynamics code SNSPH (Fryer \& Warren 2002; Warren
et al.\ 2004).  Below we discuss the details of each of these steps
individually, including a discussion describing our technique to map
each preceeding step with the next.

\subsection{Initial Star Evolution}

In this first paper, we focus on a two different helium cores: 8 and
16\,M$_\odot$ helium stars at the onset of central helium burning.  It
is likely that the more massive star has evolved beyond a pure helium
core.  Test calculations of the more evolved cores find that nuclear
burning during the merger can be important.  We will delay merger
calculations of evolved cores until we have better tested the network
in our SNSPH code (paper II).

The stars are evolved to this state as single stars using the stellar
evolution code KEPLER (Weaver et al. 1978; Woosley et al. 2002).  We
follow this single star evolution until the onset of mass transfer,
before helium ignition for our pure helium cores, at carbon ignition
for our compact star.  These stellar models are the input for our
multi-dimensional merger calculations.

\subsection{The Merger Process}

To study the merger of these stars, we must map these stars into
3-dimensional binary systems at the onset of mass transfer.  We have
three different merged systems: 8+8M$_\odot$, 16+16M$_\odot$, and
8+16M$_\odot$ binaries.  Because both stars have to evolve nearly at
the same time, the equal massed systems are the most-likely systems
for the He-merger GRB scenario that we are considering here.  If the
secondary star gains considerable mass from the primary during a first
mass transfer phase (prior or instead of the common envelope pictured
in Fig. 1), the secondary helium core could actually be more massive
than the primary helium core.  The merger of a 8+16M$_\odot$ system
represents an extreme case of this scenario.  We map the stars into
shells of smooth particle hydrodynamics (SPH) particles.  For this
multi-dimensional evolution, we use a simplified version of the
parallel SNSPH code (Fryer \& Warren 2002, Warren et al.\ 2004).  This
simplified code using a polpytropic equation of state (assuming ideal
gas: $\gamma=5/3$) and neglects nuclear burning.  Because the merger
process is so rapid (a few orbit cycles) and the particle temperatures
do not increase significantly during the merger, our assumption that
nuclear burning is unimportant holds.  Both our mapping and our
simplified equation of state lead to initial oscillations in the star.
We first model these stars as single systems, adiabatically damping
the oscillations until the star is stable.

After the stars have stabilized, we map them into a binary system.  To
test our angular momentum conservation and stability, we have run
these stars for over 10 orbits in systems roughly 2-3 times beyond
their Roche-overflow separation.  The total angular momentum was
conserved to better than 1 part in a million at the end of this
simulation and the stars remained stable with no further oscillations
(see Warren et al.\ 2004 for details).

For our actual merger calculations, we first assume that we can ignore
the hydrogen envelope aside from its viscous forces that drive the
merger.  The friction caused by the hydrogen envelope is driving these
two cores together.  Our initial binary is set with the two cores just
beyond their Roche overflow separations, but with slightly decreased
angular momenta (corresponding to the additional angular momentum that
will be lost to friction with the hydrogen envelope).  These angular
momenta are given in Table 1.  In this manner, we can mimic the
effects of the hydrogen common envelope while concentrating our
resolution on the helium cores themselves.  We assume that much of the
hydrogen envelope has been ejected in the hydrogen common envelope
phase, and the rest is ejected during the core merger or due to winds
during subsequent evolution without affecting the structure of the
core.  Given the low binding energy of the hydrogen envelope with
respect to the core, this assumption is valid \footnote{The main 
effect of any residual hydrogen envelope is to reduce the total 
mass lost due to winds.  We mimic this effect by altering the mass 
loss in the subsequent evolution of the star.}.

Within a few orbit timescales, the cores merge (Figs. 3,4).  Figure 3
shows density isosurfaces of the merger of two 16M$_\odot$ helium
stars $1.1\times 10^4,3.3 \times 10^4$s after the start of the
simulation.  Note that as the cores merge, an excretion disk is
ejected along the orbital plane.  This disk will sweep up any
remaining hydrogen envelope and continue to expand as the star
evolves.  Indeed, as the star evolves, it will lose mass in a wind,
and the excretion disk will slowly accelerate throughout the last
phase of the star's life.  Note also that most of the mass ejected in
this merger flows out in the orbital plane, not along the orbital axis
(and hence rotation axis) where the collapsar jet is likely to lie.
Figure 3 shows density isosurfaces for the merger of the 8 and
16M$_\odot$ helium stars, 5100,8100s after the start of the
simulation.  The ejecta out of this merger is much less symmetric.
These merged systems must then be mapped back into our stellar
evolution code to follow their subsequent evolution to collapse.

\subsection{Final Evolution}
\label{sec:starfinal}

Mapping the multi-dimensional simulations back into 1-dimension proved
much more difficult.  In the simplest mapping conserving mass, energy
and angular momentum, we choose a radial grid and sum the relevant
quantities for all the particles in each radial zone.  This mapping of
a multi-dimensional simulation into 1-dimensional code with different
equations of state conserves energy but not pressure gradients,
however.  Not surprisingly, the new 1-dimensional star was not in
equilibrium.  We adjusted the temperature such that for the given
density structure, which we preserved, a hydrostatic model resulted.
We then simulated the remaining evolution till onset of core collapse
using the KEPLER code including angular momentum transport and
rotationally induced mixing (Heger et al.\ 2000).

After the mapping step the star relaxes to thermal equilibrium on a
Kelvin-Helmholtz time scale.  At this point, another difficulty of the
mapping became apparent: The outer layers of the relaxed model were
vastly super-Keplerian when contracting to a non-rotating (not
considering centrifugal forces) thermal equilibrium density structure.
We tested different techniques to remove this unphysical situation.
The three most prominent are: 1) removing the entire super-Keplerian
outer layer after the stars has reached thermal equilibrium (removing
the material at an earlier stage of higher density could have caused
excessive mass loss); 2) rotationally induced enhancement of the mass
loss rate.  In particular, as long as the star exceeds the
``Omega-limit'' (Langer 1997), high mass loss rates could occur,
formally diverging in Langer's prescription, but here we limited the
mass loss to 100$\times$ the normal non-rotational mass loss rate.
This ensures that the outer super-Keplerian layers are lost on a
time-scale short compared to the evolutionary time-scale but at the
same time allows the star to adjust its structure to the mass
reduction.  3) remove super-Keplerian angular momentum in the outer
layers without removing mass.  The physical motivation for this mode
of angular momentum loss is the possibility that an excretion disk
could form around that star and transport angular momentum away while
removing only very little mass. (The general idea behind disks, though
accretion disks in most pictures studied, is that they transport mass
in while transporting angular momentum outward.)  In this paper, we
focus on the results using assumption 1: Series ``a'', assumption 2:
Series ``b'', and assumption 3 without any mass loss: Series ``c''
(see Table 1).
  
Our different techniques of treating the super-Keplerian outer layers
after the remapping produce different stellar structures at core
collapse due to different amounts of mass loss and different
rotationally induced mixing.  These differences reflect part of the
errors in these calculations and we will review the results from both
in our analysis (Table 1).  In addition, the mass-loss from winds is
uncertain.  For our fast rotating merged systems, this mass-loss is
even less known.  The effects of rotation on mass-loss is just now
being studied.  In addition, we have assumed that the hydrogen
envelope is ejected during the merger.  But this assumption may not be
true.  The existence of a hydrogen envelope will not effect the
hydrodynamic merger calculations, but this hydrogen envelope must be
blown off before the He-core will lose mass, effectively delaying the
Wolf-Rayet phase.  Hence, we have also run some of our stars assuming
no mass loss from the system after the binary merger (Series ``c'' -
see Table 1).  Such a simulation mimics what we might expect if the
hydrogen envelope is not entirely ejected during the merger process.

\section{Simulation Results}

From Figures 3 and 4, we see that much of the initial mass and angular
momentum is lost in the ejecta from the merger.  Even so, the merged
cores still retain 1-2 orders of magnitude more angular momentum than
necessary to form an accretion disk around the nascent black hole.
Figure 5 shows the equatorial angular momentum profile of all our
models, both right after the merger and at collapse (after its final
evolution).  Note that most cores lose 99\% of their angular momentum
in the evolution after the merger.  Without mass loss, the angular
momentum is a factor of 3-10 times greater.  This is because it is the
mass loss from winds that carries away much of the angular momentum.

Also plotted on Figure 5 is the angular momentum necessary for the
infalling material to hang up at the Scharzschild radius in a
centrifugally supported disk ($j=\sqrt{2 GMr_{\rm Sc}}$ where $r_{\rm
Sc}=3 GM/c^2$, $G$ is the gravitational constant, $M$, the enclosed
mass, and $c$, the speed of light).  If the angular momentum of the
star roughly exceeds this value, a disk will form during
collapse\footnote{Note that if the black hole has angular momentum,
this minimum angular momentum value will lower, to as little as
$1/\sqrt6$ for a maximally rotating black hole.  On the other hand,
for any engine to work, the disk must form beyond the innermost stable
circular orbit.  How large a disk depends on the still uncertain
collapsar engine.}.  The mass position where this occurs gives us the
rough mass of the black hole at disk formation.  We see that although
some shells of material below 3M$_\odot$ have sufficient angular
momentum to hang up in a disk, it is not until beyond 3M$_\odot$
(4M$_\odot$ for the he1616 and he816 models) that the infalling
material will consistently hang up in a disk of material for our
models with mass loss.  Hence, for most of our systems, we expect
black hole masses for the collapsar engine above 3M$_\odot$.

To determine the true fate of the star (e.g., whether it forms a black
hole or neutron star; if, and when, a collapsar jet is launched,
etc.), we must also study the density and entropy profiles of these
stars.  Figure 6 shows the entropy versus enclosed mass for all our
models.  It is the entropy in the core that determines the strength of
the bounce when the core collapses down to a proto-neutron star.  In
general, the entropy of the Series ``c'' models (no mass loss) is
higher in the inner 1\,M$_\odot$ core and lower just beyond that core,
but these differences are neither absolute or very strong.  Figure 7
shows the density profiles for these same models.  Those compact stars
where the density remains high to masses beyond 2M$_\odot$ (namely the
models with no mass loss - Series ``c'') are more difficult to explode
and will more-likely collapse to form black holes.  We discuss this in
more detail in section \ref{sec:fate}.

Lastly, we show the nuclear abundances at collapse of 4 of our models
which give the full range of possible composition structures for these
merged cores (Fig. 8).  The smallest iron core ($\sim 1.8$M$_\odot$)
is produced by the merger of the two smallest stars (He88), but note
that the iron cores (and silicon layers) for the He816a and and
He1616a models are nearly identical ($\sim 2$M$_\odot$).  This is
because there is such mass loss in the He1616a model that its core
reflects that of a much smaller star.  With mass loss turned off
(He1616c), the iron core for the 16+16M$_\odot$ models is even larger
($\sim 2$M$_\odot$) and the C-free (oxygen burning) layer extends to
6M$_\odot$ (compared to $\sim 1.8$M$_\odot$ for the He88a model and
$\sim 2.5$M$_\odot$ for the He816a,He1616a models).  This large
silicon shell will ultimately play a major role in the fate of these
stars.

\section{The Fate of Merged Cores}
\label{sec:fate}

From the results of our simulations, we can now estimate i) if the
collapse of such stars will produce a neutron star or black hole, ii)
assuming a black hole is formed, if, and when, an accretion disk will
form and iii) the delay between collapse and jet formation.  As we go
through each of these steps, we study an increasingly select set of
merged systems, ultimately determining the ideal set of collapsar
candidates.  We can then use the conditions required to make these
candidates as a constraint on the population of gamma-ray burst
progenitors (Section \ref{sec:pop}).

The success or failure of core-collapse supernovae remains an unsolved
problem in astrophysics (Herant et al. 1994; Burrows, Hayes, \&
Fryxell 1995; Janka \& M\"uller 1996; Mezzacappa et al.\ 1998; Fryer
1999; Buras et al.\ 2003), and core-collapse theorists have not yet
identified all of the factors that determine the fate of a
progenitor star.  However, Fryer (1999) has found that the density
profile is a good indicator on whether an explosion can occur or
not.  He argued that the sharp drop in density for the collapsed core
of stars between 8-20M$_\odot$ is what allows the convective region
above the proto-neutron star to expand and drive a supernova
explosion.  Fryer (1999) used the accretion rates on the convective
core as a function of time after core bounce as a diagnostic of the
fate of massive stars.  This accretion exerts a ram pressure that must
be overcome to drive an explosion.  Low accretion rates mean a weaker
ram pressure which is easier to explode.  The corresponding accretion
rates for our merger models are plotted in Figure 9.  The rates for
the 15M$_\odot$ and 25M$_\odot$ stars are plotted for comparison.  

Many of the merger models with mass-loss in this paper have very
similar accretion rates to the 15M$_\odot$ star.  In Fryer (1999), the
15M$_\odot$ exploded quickly (strong explosion, neutron star compact
remnant).  Observational evidence also suggests that 15M$_\odot$ stars
do produce strong supernova explosions and neutron star remnants.  It
is likely that merger models that have accretion profiles similar to
the 15M$_\odot$ star (He88a, He88b, He1616b, He1616b) also produce
strong explosions with neutron star remnants.  Fryer (1999) found that
the 25M$_\odot$ star took longer to explode, producing a weak
explosion and a black hole forming from fallback.  The remaining
models are much more similar to the 25M$_\odot$ star and it is likely
that these models form black holes either directly (collapsar
Type I) or through fallback (collapsar Type II).

Our merger models, by ansatz, satisfy Criterion III (loss of hydrogen
envelope - \S \ref{sec:intro}) of GRB progenitors.  In the last
paragraph, we found that He816a, He1616a, and ``c'' series all seem to
satisfy Criterion I (collapse to black hole).  The final criterion
that must be satisfied is the angular momentum constraint: the
infalling material must have enough angular momentum to form a disk
around the black hole.  Figure 10 compares the angular momenta from
the black-hole forming mergers with massive single stars.  From Figure
5, we see that there is sufficient angular momentum to form a disk in
all our black-hole forming models.  Models he1616a and he1616b do not
have dramatically more angular momentum than massive single stars.
Those models without mass loss (``c'' series), however, collapse with
angular momenta 3-10 times higher than single star models.  Recall
that these zero mass-loss models mimic the situation where a hydrogen
envelope persists around the helium core after the merger or those
evolved systems that have little mass-loss before collaspe.  It is
likely that this hydrogen will be removed prior to the collapse of the
core.  But it could delay the Wolf-Rayet stage long enough to minimize
the effects of mass-loss.

Another way to reduce the post-merger mass-loss is to assume the primary 
system has evolved.  If the core has already undergone some helium 
burning, the time between merger and collapse will be shorter.  This 
will both reduce the mass-loss and hopefully reduce the amount of 
angular momentum lost through viscous forces.  We expect such systems 
to rotate faster than those of our current study.

The angular momentum in these black-hole forming stars is so high that
the black hole formed in these collapses is close to be maximal
rotation.  This formation process begins with the collapse of the core
to a rapidly spinning proto-neutron star.  The fast rotating core
(Fig. 5) collapses to a proto-neutron star spinning nearly at
break-up.  This support leads to a more massive maximum neutron star
mass.  But because the angular momentum increases with radius, the
increase in the maximum mass will not be much more than the 20\%
expected for a uniform rotator (Cook et al.\ 1994; Morrison et al.\
2004).  We expect the black hole to form from a roughly 3M$_\odot$
neutron star with enough angular momentum to produce a maximally
rotating black hole.  Such high angular momentum will aid both the
magnetic field and neutrino annihilation engines driving collapsar
jets.

For magnetically-driven collapsars, the launch of the jet can occur as
soon as an accretion disk is formed.  For example, Katz 1997 argues
that it takes only a few differential revolutions of the disk to build
up the magnetic field strength necessary to produce a jet.  With the
large amount of angular momentum in the black hole and the disk, the
magnetically-driven jet model has a large reservoir of rotational
energy to drive an explosion.  In this scenario, the delay between the
GRB jet and the initial neutrino/gravitational wave signal can be very
short (limited to the time it takes for the proto-neutron star to
collapse to a black hole $<$1\,s).

Neutrino annihilation is a different matter.  The efficiency of energy
deposition via neutrino annihilation is very low.  Hence, neutrino
annihilation will not drive a jet until sufficient accretion along the
axis of rotation has occurred to clear out a funnel.  As the axis
clears, the energy deposition will be sufficient to drive an explosion
(Fryer \& Meszaros 2003).  Fryer \& Meszaros (2003) showed that the
black hole mass at, and the delay from collapse to, black hole
formation can be solved semi-analytically for a given stellar density
profile.  Figure 11 shows the density profiles for our merged cores in
comparison to single star models.  From these densities, we derive the
accretion rate along the rotation axis as a function of time
(Fig.~12).  When this rate falls below the critical rates derived by
Fryer \& Meszaros (2003), the jet is launched\footnote{For a given
neutrino energy deposition (which depends on viscosity in the disk and
accretion rate through the disk), exists a critical density below
which a neutrino-driven jet is launched.  This critical density has a
corresponding critical accretion rate along the rotation axis.}.  Note
that these estimates all predict black hole masses above
$\sim$10M$_\odot$ and delay times above 50s (Table 1).  It is unlikely
that the neutrino driven mechanism will work for these stars.  The
neutrino driven mechanism may not work for any stars that will
ultimately form black holes.

\section{Implications for Neutron Star, Black Hole, and GRB Populations}

Close binaries are observed.  Systems that don't merge produce X-ray
binaries and double neutron star binaries such as the Hulse-Taylor
pulsar system.  Those that do merge may produce equally interesting
objects, from SN 1987A to GRBs.  This paper focuses on the evolution
of a class of progenitors for GRBs known as He-star/He-star-mergers
(Fryer et al. 1999).  We find that, under conditions with little
mass-loss, He-star/He-star-mergers systems have 3-10 times more
angular momentum than single star collapsar progenitors.
Nevertheless, if the mass loss is high, these merged systems form
neutron stars, not black holes.  Unfortunately, the true fate of any
binary is extremely sensitive to mass loss, and our capability to
predict this, in turn, is limited by our knowledge of, or lack
thereof, wind mass loss in rotating stars as well as the amount of
hydrogen left in these merged systems which depends upon the specifics
of the merger event itself.  Even with these uncertainties, we can
already use our current results to predict the neutron stars, black
holes and GRBs formed by these binaries.  Before we do, let's review
once more the mass loss uncertainties.

\subsection{Mass-Loss Uncertainties}

The subject of mass loss from winds has long been in dispute between
the stellar and X-ray binary communities.  The stellar community has
long argued that both observations of stars and the number of
Wolf-Rayet stars require high wind mass loss rates.  The smooth
mass-loss prescriptions produced for stellar theorists generally cause
most massive stars to lose most of their mass prior to collapse.
Solar metallicity stars in binaries that lose their mass during Case B
mass transfer (during the expansion off of the main sequence) have no
chance of forming black holes without artificially lowering the
currently predicted mass-loss rates (see Fryer et al.\ 2002 for a
review).  And yet Case B mass-transfer binaries dominate the
production of X-ray binaries and X-ray binaries are observed in
considerable numbers in the Galactic disk.  

The situation is such that either a) mass-loss rates are lower (which
stellar theorists argue can not be the case because then they can not
make Wolf-Rayet stars) or b) binary population synthesis theorists
have not found the correct path for making X-ray binaries.  Solving
this problem is essential to understanding the progenitors of GRBs and
requires a better physical understanding of mass loss (which may
include pulsational studies - Joyce Guzik, private communication), we
believe lower mass loss rates, at least for some stars, will be the
ultimate solution.  This is, in part, because of the false assumption
that the numbers of Wolf-Rayet stars require high mass-loss rates.
Recall (Fig. 2) that binaries can also make Wolf-Rayet stars.  Indeed,
Podsiadlowski et al. (1992) have shown that binaries can dominate the
Wolf-Rayet rate.  On the other hand, Meynet \& Maeder (2003) argue
that the fractions of Wolf-Rayet stars as a function of metallicity is
better fit by single star models than binary systems, implying that
single stars are the dominant Wolf-Rayet progenitor.  So the
progenitors of Wolf-Rayet stars remains a matter of contention.

There are additional uncertainties in binary merger calculations.  In
this paper, it was assumed that the hydrogen envelope provided the
last bit of friction to merge the binaries, but was otherwise ignored.
It could be that even after the merger, some hydrogen remained on top
of the star.  Although this does not affect the hydrodynamical merger
process significantly, it will change the mass-loss dramatically as
the star will not enter the Wolf-Rayet phase until after this hydrogen
is shed.  Unfortunately, the amount of hydrogen remaining will depend
upon the binary characteristics.  It is likely that our studies, in
the near future, will be limited to qualitative trends.  But even with
these qualitative trends, we can say much about the resulting compact
systems and explosions from these mergers.

\subsection{Neutron Star Remnants}

Those systems with small helium cores (either through smaller mergers
not studied in this paper or those that lose considerable mass through
winds) will form rapidly rotating neutron stars, with rotational
velocities at least as large as the fastest rotating single stars.
These stars will collapse to form proto-neutron stars surrounded by a
disk (thermally and centrifugally supported) that is likely to
dissipate the angular momentum before the material contracts and gains
considerable kinetic energy (Fryer \& Heger 2000; Fryer \& Warren
2004).  If large magnetic fields are produced in these proto-neutron
stars, this angular momentum dissipation will occur even earlier.
With our current understanding of the disk evolution, these objects
will {\it not} produce GRBs (which require periods below 1 ms), but they
will form fast-spinning pulsars.

Because the mass loss increases with metallicity, we should see more
neutron star systems with increasing mass.  Since the neutron star
population will be dominated by smaller helium mergers, this increase
may easily be hidden.  The black hole formation rate, on the other 
hand will change dramatically.

\subsection{Black Hole Remnants}

Those mergers that don't lose considerable mass will collapse to form
black holes.  Because of the low mass-loss, these stars have 3-10
times more angular momentum just prior to collapse.  The black holes
will be born, preferentially at low metallicities (depending on our
understanding of mass loss) with spin rates that are higher than those
produced in single stars.  The high angular momentum in the infalling
star will form a disk around the black hole, possibly forming a GRB
jet.

\subsection{GRBs}
\label{sec:pop}

The collapsar engine for Gamma-ray bursts has three possible
progenitors (Fryer et al.\ 1999): single stars, the merger of a
compact remnant and its binary companion (He-merger), and the merger
of two helium cores (He-star/He-star merger) studied in this paper.
Single star progenitors suffer from having too little angular
momentum.  As we have seen in this paper, this angular momentum
problem may be somewhat alleviated with He-star/He-star mergers.  The
He-merger model could possibly have the opposite problem: too much
angular momentum (see Zhang \& Fryer 2001, Di Matteo et al.\ 2002).

All three of these progenitors may be contributors to the GRB
population, but they have very different redshift distributions.
Both the helium-merger and He-star/He-star merger scenarios can occur
at large redshifts (low metallicities).  The single star progenitors
decrease significantly with lower metallicity.  If they are the
sole contributor, there should be very few long-duration bursts at
high redshifts and none from population III stars even though the
initial mass function might be skewed toward massive stars at these
high redshifts.  This sharp decrease is because single stars can not
blow off their hydrogen envelopes without metals.  It is harder to
determine the redshift evolution of binary systems without detailed
population synthesis calculations.  Lower mass-loss and the skewed
initial mass function increase the number of close binaries that
collapse to black holes.  However, low and zero metallicity stars tend
to expand less in their giant phases, leading to fewer mass transfer
phases.  Keeping in mind these uncertainties, we still predict the
fraction of stars that form GRBs under binary scenarios increases with
redshift.  The redshift distribution of GRBs will allow us to easily 
distinguish between GRB progenitors.

If He-star/He-star mergers are the dominant progenitor of GRBs, we can
make a few further predictions beyond the increase in GRB formation
rate with redshift.  The progenitors are more massive and have more
angular momentum with increasing redshift.  Hence, there should be a
steady evolution of the GRB energy with increasing redshift.  Although
it is certainly plausible for these more massive, faster-rotating
stars to make stronger bursts at higher redshifts, without a full
understanding of the GRB engine, no reliable predictions can be made.

\acknowledgements 
This work was funded under the auspices of the U.S.\ Dept.\ of Energy,
and supported by its contract W-7405-ENG-36 to Los Alamos National
Laboratory, by a DOE SciDAC grant number DE-FC02-01ER41176 and by NASA
Grant SWIF03-0047-0037.  The simulations were conducted on the Space
Simulator at Los Alamos National Laboratory.

\clearpage
\begin{deluxetable}{lcccc}
  \tablewidth{38pc} 
\tablecaption{Summary of Simulations}

\tablehead{ \colhead{Model} & \colhead{Binary Masses} 
& \colhead{Ang.~Mom.} & \colhead{KEPLER\tablenotemark{a}} 
& \colhead{Remnant}\\
  \colhead{Name} & \colhead{(M$_\odot$)}  
& \colhead{($10^{57}\,{\rm erg\,\, s}$)} 
& \colhead{mapping} & \colhead{Fate}}

\startdata

M88a   & 8+8   & 2.4 & 1 & NS \\
M88b   & 8+8   & 2.4 & 2 & NS \\
M88c   & 8+8   & 2.4 & 3-no wind & BH \\
M1616a & 16+16 & 7.3 & 1 & BH \\
M1616b & 16+16 & 7.3 & 2 & NS \\
M1616c & 16+16 & 7.3 & 3-no wind & BH \\
M816a  & 8+16  & 2.3 & 1 & BH \\
M816b  & 8+16  & 2.3 & 2 & NS \\
M816c  & 8+16  & 2.3 & 3-no wind & BH \\

\enddata

\tablenotetext{a}{The methods used to map from the SNSPH calculations
  to KEPLER are described in \S \ref{sec:starfinal}.  Methods 1, 2,
  and 3 are described in detail in that section.  Briefly, 1
  corresponds to removing the super-Keplerian mass, 2 corresponds to
  using an enhanced rotational mass-loss, and 3 corresponds to
  reducing the angular momentum of the super-Keplerian material
  without removing the mass.  In Series ``c'' we turn off mass-loss
  for the entire stellar life.}

\end{deluxetable}

\clearpage
\begin{figure}
\vskip -50pt
\epsscale{.80}
\plotone{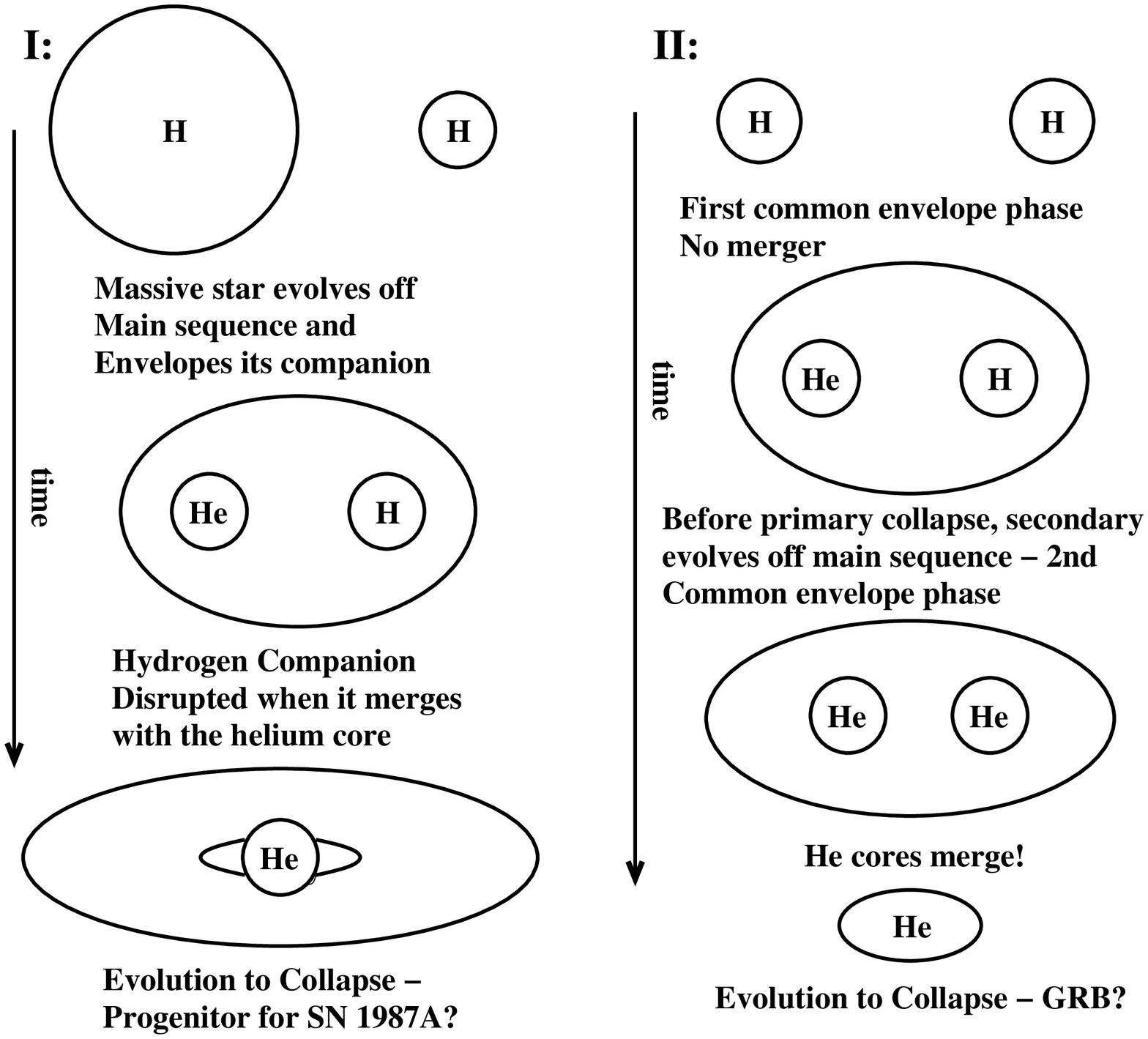}
\vskip -50pt

\caption{Two evolutionary paths for close binaries:  I) a close binary 
  system where the masses of the two stars differ by more the 10\%,
  II) a close binary system where the masses of the two binary stars
  are nearly identical.  In system I, the more massive star evolves
  off the main sequence and expands to envelope its companion.
  Friction causes the orbit in this system two decay.  In some cases,
  the orbit will decay so much that the systems merge.  The less
  massive hydrogen star is disrupted when it reaches the dense helium
  core of the more massive star.  Podsiadlowski (1992) argued that SN
  1987A formed in such a scenario.  System II involves two stars of
  nearly identical mass.  In the first common envelope phase, the
  system tightens, but does not merge.  However, the similar masses of
  the two stars mean that the less massive star evolves off the main
  sequence before the more massive star collapses, leading to a second
  common envelope phase and the merger of the two helium cores.  Fryer
  et al. (1999) proposed this scenario as a formation scenario for
  collapsars.}

\end{figure}
\clearpage

\begin{figure}
\vskip -50pt
\epsscale{.80}
\plotone{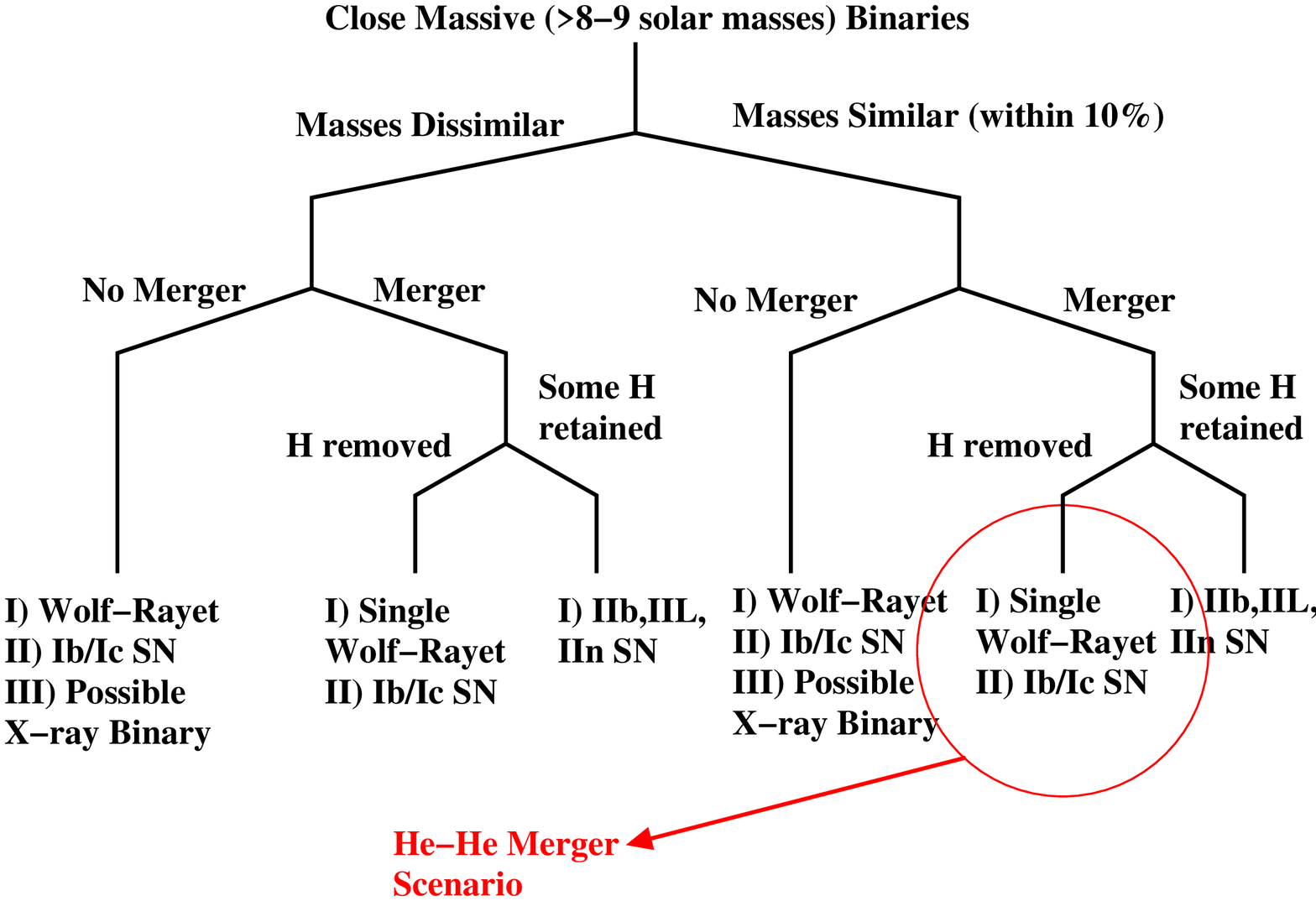}
\vskip -120pt
\caption{A diagram of the fates of close, massive binaries.  
  X-ray binaries are formed in systems where the two stars do not
  merge.  The traditional path for the formation of X-ray binaries
  would be a system with dissimilar stellar masses (masses dissimilar
  route).  The common envelope phase removes the hydrogen envelope of
  the more massive star, turning it into a Wolf-Rayet star, but the
  system does not merge (no merger route).  If the system remains
  bound after collapse, an X-ray binary is born.  Many more (the exact
  number is sensitive to the population synthesis assumptions - see
  Fryer et al. 1998) of these dissimilar binaries lead to the ultimate
  merger of the two stars.  The increased mass-loss due to this merger
  may be enough to allow a star, that would otherwise not become a
  Wolf-Rayet star, to lose its entire hydrogen envelope.  Hence, these
  binary systems produce single Wolf-Rayet stars.  We are more
  interested in those binaries that consist of stars that are very
  close in mass.  The case where these stars do not merge is now
  considered to be one of the primary formation scenarios of double
  neutron star systems like the Hulse-Taylor pulsar system (Brown
  1995; Fryer et al. 1999; Belczynski et al. 2002).  Those systems
  that merge and ultimately lose their hydrogen envelope may be
  collapsar progenitors, the so-called He-star/He-star merger scenario
  (Fryer et al. 1999).}
\end{figure}
\clearpage

\begin{figure}
\plotone{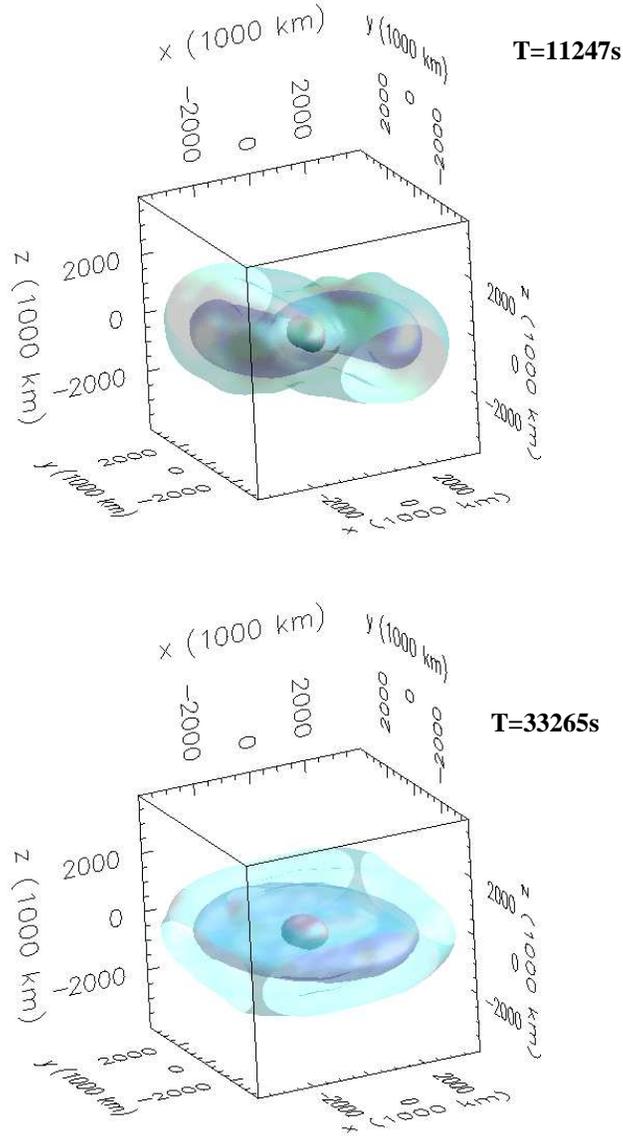}
\caption{Density contours at 2 different times of the merger of 
  2 16M$_\odot$ stars.  The contours correspond to (from inner to outer
  contour) densities of $8\times 10^{-2},10^{-4},10^{-5} {\rm g \, cm^{-3}}$
  respectively.  For the mergers of identical mass stars, the matter
  is ejected in an axis-symmetric excretion disk.}
\end{figure}
\clearpage

\begin{figure}
\plotone{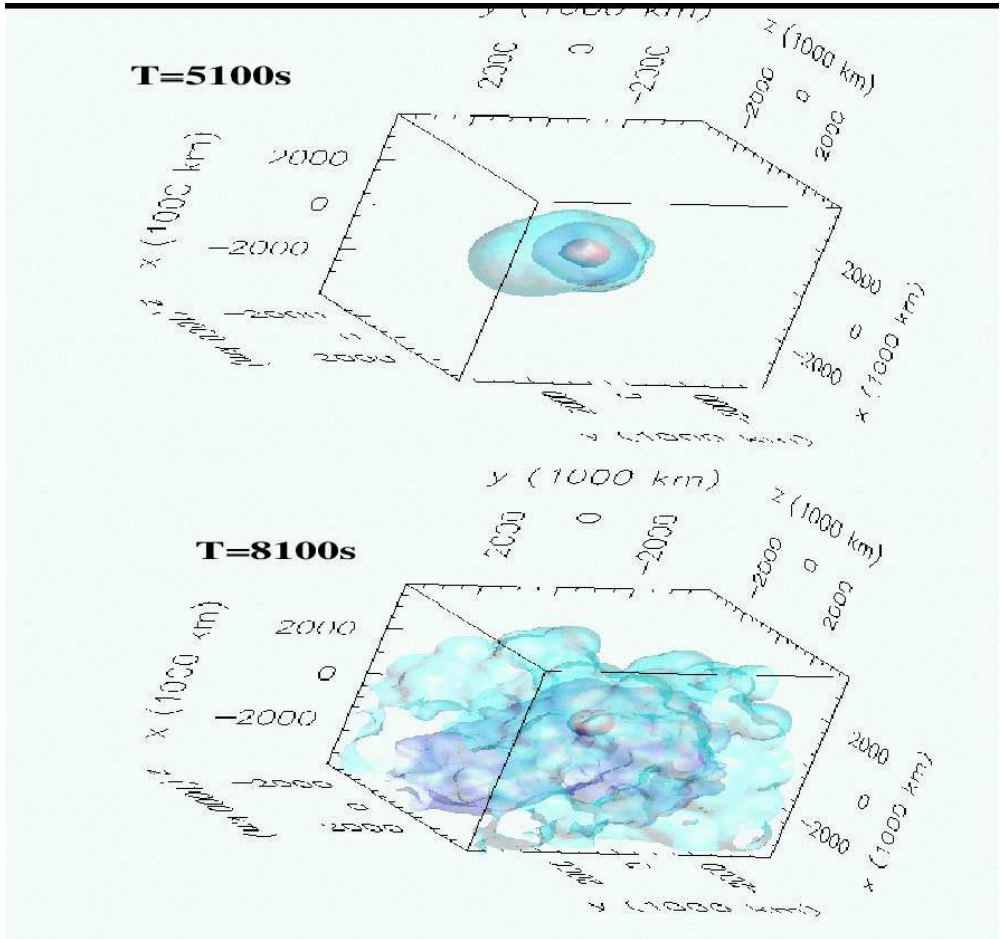}
\caption{Density contours at 2 different times of the merger of 
  an 8M$_\odot$ star with a 16M$_\odot$ star.  The contours correspond to
  (from inner to outer contour) densities of
  $8\times 10^{-2},10^{-5},10^{-9} {\rm g \, cm^{-3}}$ respectively.  The
  merger of stars with different masses leads to a much less symmetic
  mass ejection than those of equal mass stars (Fig. 3).}
\end{figure}
\clearpage

\begin{figure}
\epsscale{.80}
\plotone{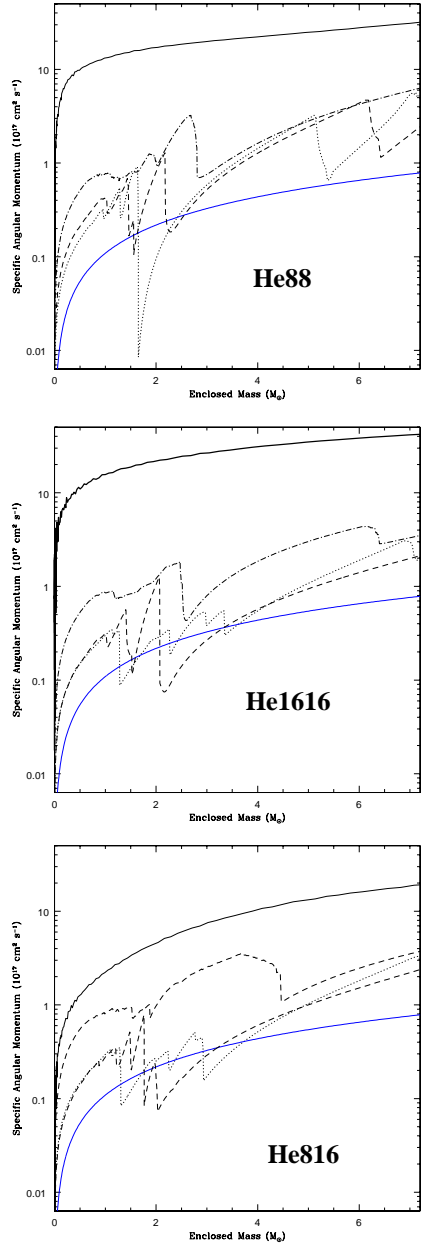}
\caption{Mean angular momenta versus mass for all
  our models just after the merger and at collapse.  The thick solid
  line at the top of the graph is the angular momentum just after
  merger.  Dotted lines correspond to ``a'' series models, dashed
  lines correspond to ``b'' series models, and dot-dashed lines
  correspond to ``c'' series models.  We've also plotted the angular
  momentum for material at innermost stable circular orbit for a
  non-rotating black hole versus mass (thin solid line).  Although the
  criterion for the formation of a disk that can drive an explosion is
  uncertain, it is likely that the angular momentum must consistently
  exceed this value. For most of our models, this is above
  3-4\,M$_\odot$. }
\end{figure}
\clearpage

\begin{figure}
\plotone{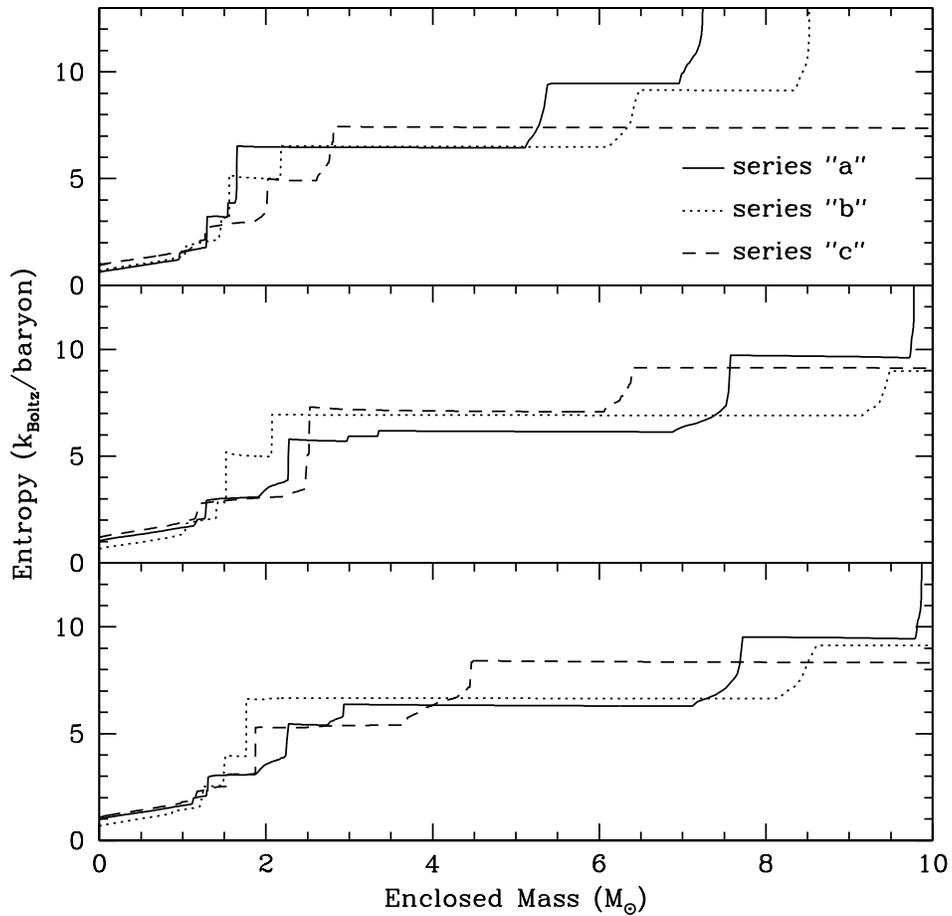}
\caption{Entropy versus mass at collapse for all models (top panel:
  he88, middle panel: he1616, bottom panel he816).  Solid lines,
  dotted lines, dashed lines correspond to Series ``a'',''b'', and
  ``c'' respectively.  Although there is considerable scatter, in
  general, the Series ``c'' models (no mass loss) have higher
  entropies in the inner 1\,M$_\odot$ core with lower entropies
  beyond.  This trend is consistent with the larger cores for these
  models.}
\end{figure}
\clearpage

\begin{figure}
\plotone{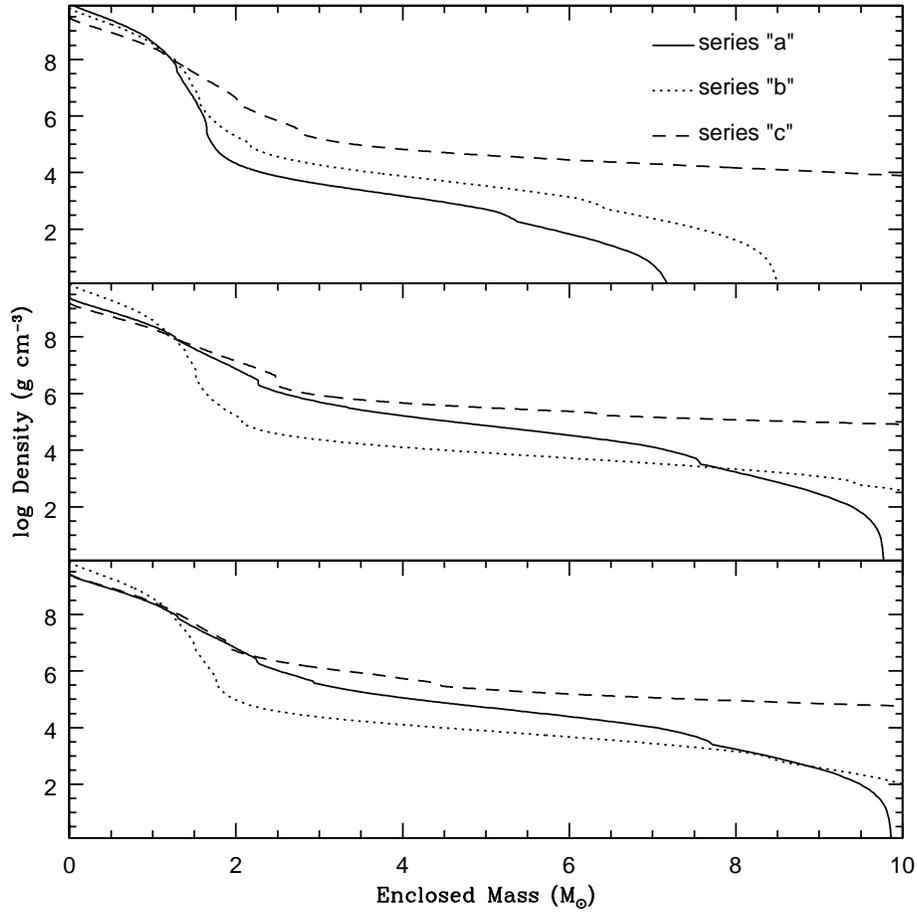}
\caption{Density versus mass at collapse for all models (top panel:
  he88, middle panel: he1616, bottom panel he816).  Solid lines,
  dotted lines, dashed lines correspond to Series ``a'',''b'', and
  ``c'' respectively.  As we could surmise from the entropy plot
  (Fig. 6), beyond $\sim 1$\,M$_\odot$, the density is higher for the
  larger cores in the Series ``c'' models.}
\end{figure}
\clearpage

\begin{figure}
\epsscale{.80}
\plotone{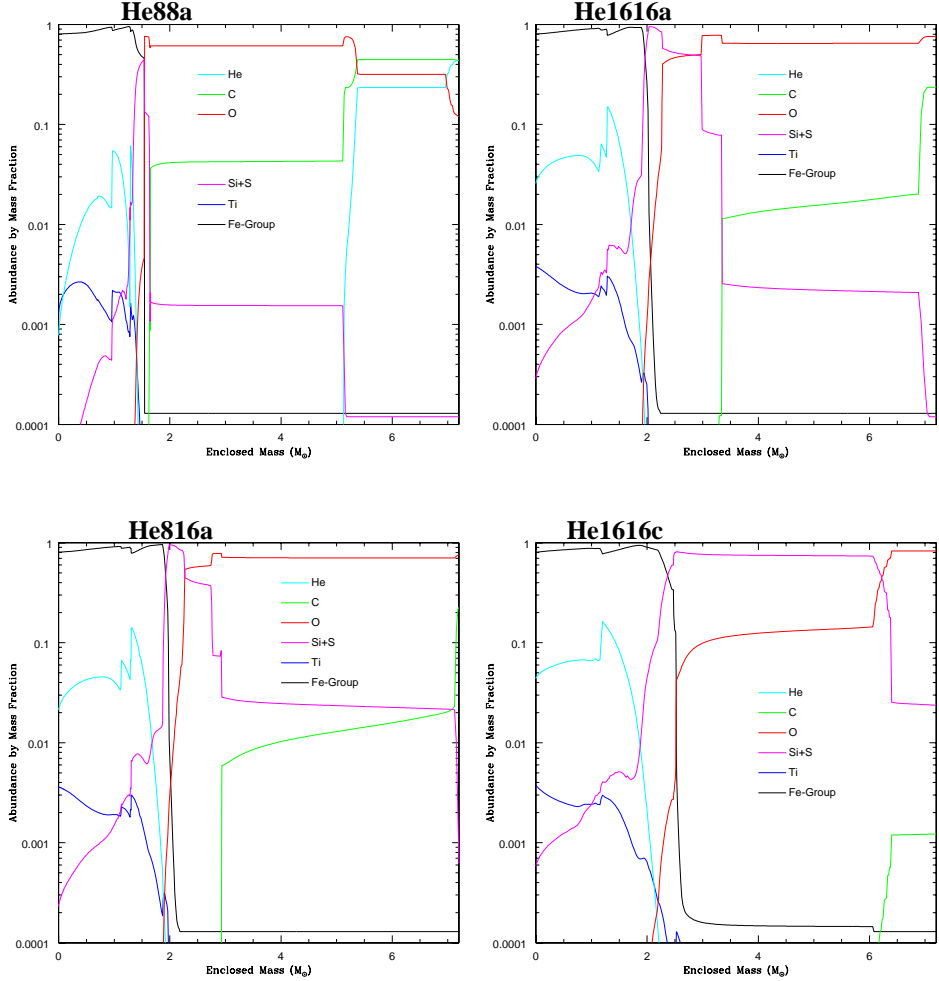}
\vskip -50pt
\caption{
  Abundance fractions versus mass for 4 merged cores just prior to
  collapse.  The smallest iron core ($\sim 1.8$M$_\odot$) is produced
  by the merger of the two smallest stars (He88).  But the trend of
  smaller cores with smaller merged objects doesn't seem to follow
  with our other models.  Mass loss causes the iron cores (and silicon
  layers) for the He816a and and He1616a models are nearly identical
  ($\sim 2$M$_\odot$).  The largest iron core arises from the merger
  of two 16+16M$_\odot$ stars without mass loss ($\sim 2$M$_\odot$).
  Even more important for black hole formation is the size of the
  silicon/sulfur layer.  This layer extends to 6M$_\odot$ for the
  He1616c model (compared to $\sim 1.8$M$_\odot$ for the He88a model
  and $\sim 2.5$M$_\odot$ for the He816a,He1616a models).  }
\end{figure}
\clearpage

\begin{figure}
\plotone{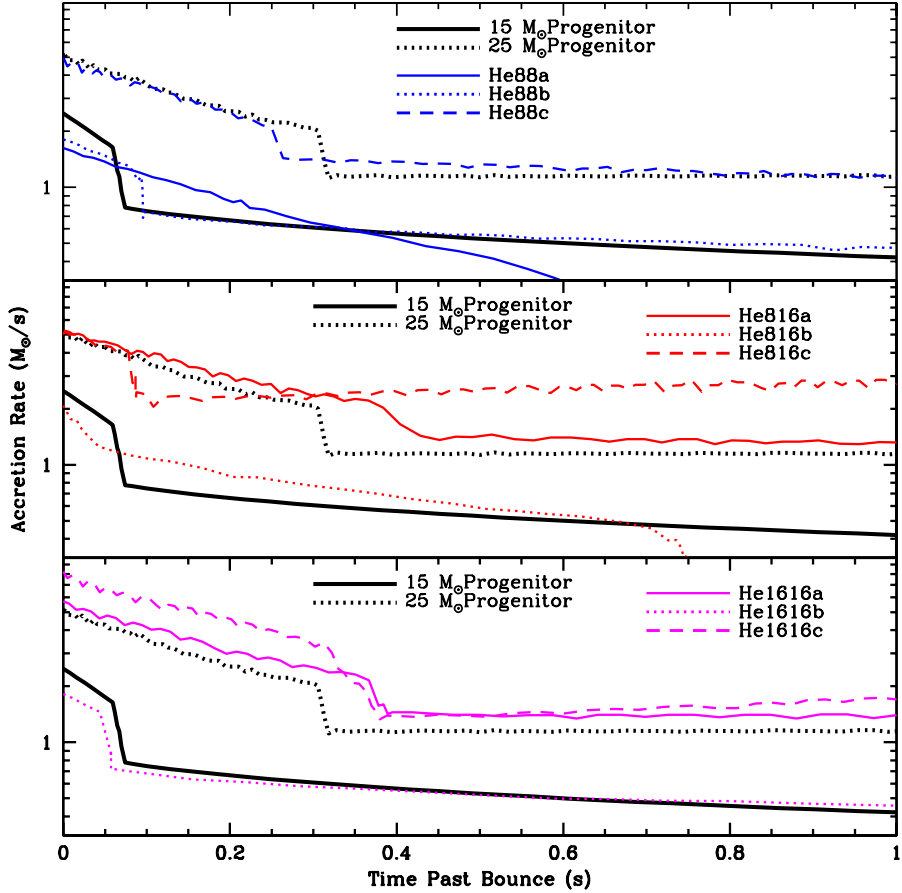}
\caption{Accretion rates onto the proto-neutron star as a function 
  of time after the bounce of the collapsed core for 15 and
  25M$_\odot$ stars (Fryer 1999) and for our new merger models.  Fryer
  (1999) found that the 15M$_\odot$ star exploded 90\,ms after
  collapse when the accretion rate dropped dramatically as the silicon
  layer (with its lower density) hit the proto-neutron star.  The
  merged stars with similar accretion rates are also likely to
  explode.  Stars the mimic the 25M$_\odot$ star are more likely to have
  weak or no explosions and collapse to form black holes
  (He1616a, He816a, Series ``c'' models).}
\end{figure}
\clearpage

\begin{figure}
\plotone{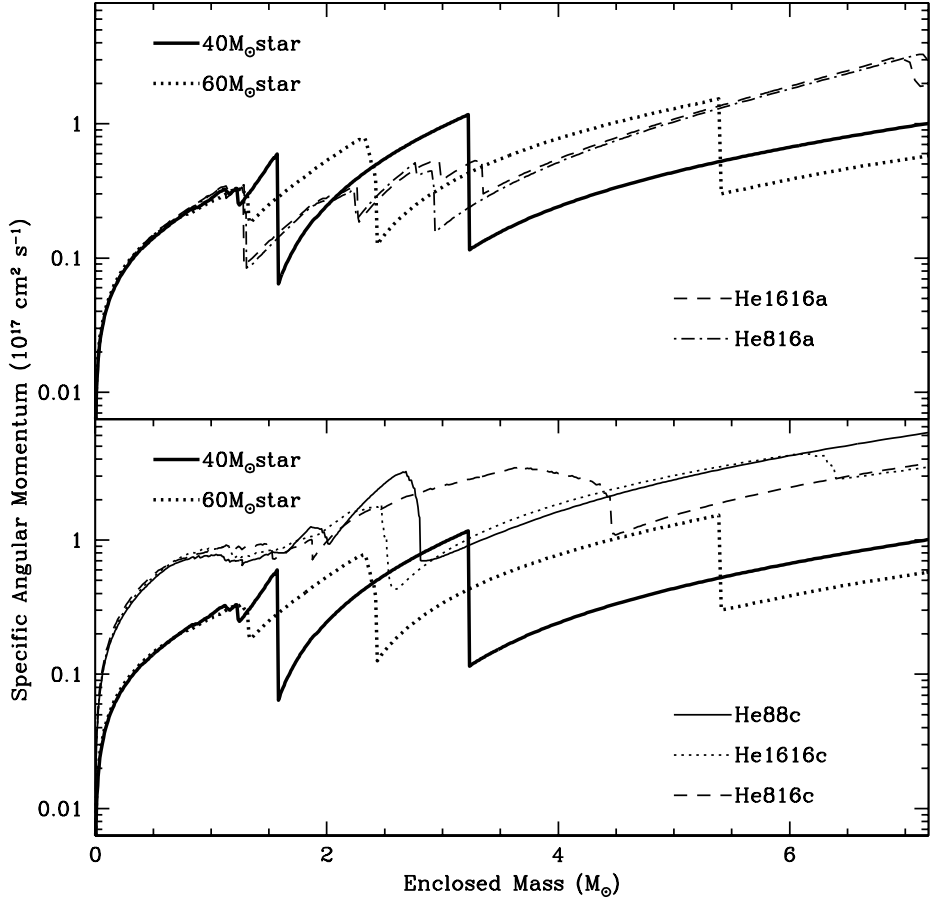}
\caption{Angular momenta versus enclosed mass for the black-hole
  forming models compared to 40,60M$_\odot$ single star models.
  He1616a and He816a (top panel) have angular momenta that are not too
  different from the single stars.  Between 2 and 3 solar masse, the
  Series ``c'' models (bottom panel), however, can have angular momenta
  that are more than a factor of 3 higher than in single-star models.
  All models, however, have angular momenta that are sufficiently high
  to produce black hole accretion disks (Fig. 5).}
\end{figure}
\clearpage

\begin{figure}
\plotone{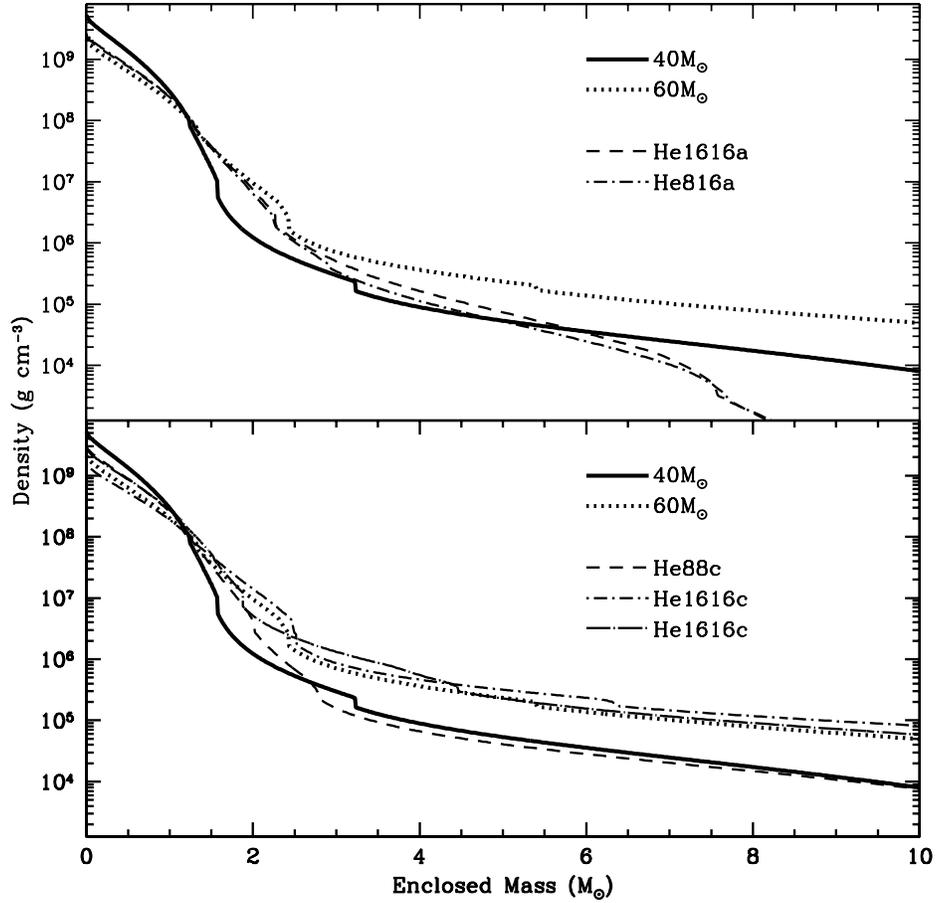}
\caption{Density versus enclosed mass for the black hole forming 
models compared to 40,60M$_\odot$ single star models.  Because 
of the high densities, similar to the single star models, 
the neutrino driven engine may fail to drive explosions 
for these models.  We can see this more clearly in Fig. 12.}
\end{figure}
\clearpage

\begin{figure}
\plotone{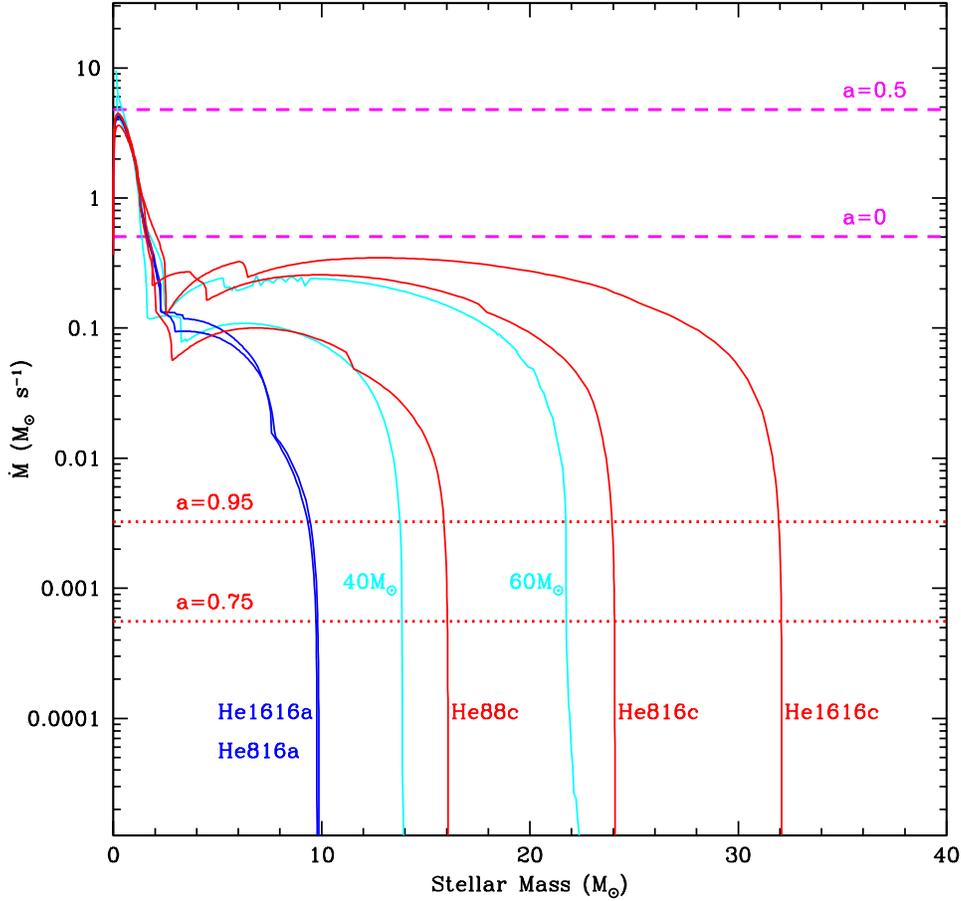}
\caption{Accretion rate along the pole versus accreted mass for all
  the black hole forming systems compared to the 40 and 60\,M$_\odot$
  single star models.  The horizontal lines correspond to disk
  accretion rates (not polar accretion) of 0.1M$_\odot$\,s$^{-1}$
  (dotted) and 1.0\,M$_\odot$\,s$^{-1}$ (dashed) for a range of black
  hole spins.  Black holes are likely to be maximally rotating, so the
  Kerr parameter $a$ is likely to reach 0.95.  Even so, for accretion
  rates of 0.1\,M$_\odot$\,s$^{-1}$, it is unlikely that the
  neutrino-driven mechanism can power jets (see Fryer \& Meszaros 2003
  for details).}
\end{figure}
\clearpage

\end{document}